\documentstyle[preprint,prl,aps]{revtex}
\begin{document}
\draft
\title{Practical solution to the Monte Carlo sign problem:\\
Realistic calculations of ${}^{54}$Fe}
\author{Y. Alhassid\cite{YAAuth}, D. J. Dean, S. E. Koonin, G. Lang,
and W. E. Ormand}
\address{W. K. Kellogg Radiation Laboratory, 106-38, California
Institute of Technology\\ Pasadena, California 91125 USA}
\date{\today}
\maketitle

\widetext
\begin{abstract}
We present a practical solution to the ``sign problem'' in the
auxiliary field Monte Carlo approach to the nuclear shell model. The
method is based on extrapolation from a continuous family of
problem-free Hamiltonians. To demonstrate the resultant ability to
treat large shell-model problems, we present results for ${}^{54}$Fe
in the full $fp$-shell basis using the Brown-Richter interaction. We
find the Gamow-Teller $\beta^+$ strength to be quenched by 58\%
relative to the single-particle estimate, in better agreement with
experiment than previous estimates based on truncated bases.
\end{abstract}

\narrowtext

Recent publications \cite{Johnson92,Lang93} have described quantum
Monte Carlo methods for exact solution of the nuclear shell model.
The methods are based on the Hubbard-Stratonovich (HS) representation
\cite{Hubbard59} of the imaginary-time many-body propagator in terms
of one-body propagators of non-interacting nucleons moving in a
fluctuating field. Thermal averages can be calculated, as can ground
state properties; errors arise only from discretization and
statistical sampling, both of which can be controlled. As these
computations scale much more gently with the number of single
particle orbits ($N_s$) and/or the number of valence nucleons ($N_v$)
than do direct diagonalization techniques, they hold great promise
for treating very large model spaces.

Unfortunately, the applicability of shell model Monte Carlo
calculations has heretofore been limited by the ``sign problem''
generic to all fermionic Monte Carlo techniques
\cite{Johnson92,Lang93,White89,Linden92}. The sign of the integrand
may vary from sample to sample and the net integral results from a
delicate cancellation that is difficult to reproduce with a finite
number of samples. The problem is well-documented (and as yet
unsolved) in simulations of correlated electron systems
\cite{White89}. Except for an important, yet schematic, class of
nuclear interactions \cite{Lang93}, we have found that all realistic
nuclear shell model Hamiltonians suffer from a sign problem.

In this Letter, we report a practical solution to the sign problem
and present the first realistic calculation of a mid-$fp$-shell
nucleus, ${}^{54}$Fe \cite{conf}. Our method is based on an
extrapolation of observables calculated for a ``nearby'' family of
Hamiltonians whose integrands have a positive sign. Success depends
crucially upon the degree of extrapolation required. We have found
that, for all of the many realistic interactions tested in the $sd$-
and $fp$-shells, the extrapolation required is modest, amounting to a
factor-of-two variation in the isovector monopole pairing strength.

A general time-reversal invariant Hamiltonian with two-body
interactions can be brought to the form
\begin{equation}
H=\sum_\alpha
\left(\epsilon^\ast_\alpha\bar{\cal O}_\alpha+
\epsilon_\alpha{\cal O}_\alpha\right)+
{1\over2}\sum_\alpha V_\alpha
\left\{{\cal O}_\alpha, \bar{\cal O}_\alpha\right\}\;,
\end{equation}
where the ${\cal O}_\alpha$ are a convenient set of one-body
operators and $\bar{\cal O}$ denotes the time-reverse of ${\cal O}$.
For real $V_\alpha$, $H$ in Eq.~(1) is a manifestly time-reversal
invariant. The auxiliary field Monte Carlo approach utilizes the HS
representation of the imaginary-time many-body propagator
$U=\exp(-\beta H)$ as a path integral over one-body propagators in
fluctuating auxiliary fields. Upon introducing $N_t$ time slices of
duration $\Delta\beta=\beta/N_t$ and complex $c$-number auxiliary
fields $\sigma_{\alpha n}$ ($n=1,\ldots,N_t$), we can write the
canonical expectation value of an observable ${\cal O}$ as
\begin{equation}
\langle{\cal O}\rangle\equiv
{{\rm Tr}\,({\cal O}e^{-\beta H})\over
{\rm Tr}\,(e^{-\beta H})}\approx
{\int D[\sigma]W(\sigma)\Phi(\sigma)
\langle{\cal O}\rangle_\sigma\over
\int D[\sigma]W(\sigma)\Phi(\sigma)}\;.
\end{equation}
Here, the approximation becomes exact as $N_t\rightarrow\infty$ and
the metric is $D[\sigma]=\Pi_{\alpha,n}\left[d\sigma_{\alpha
n}d\sigma^\ast_{\alpha n} \Delta\beta\vert
V_\alpha\vert/2\pi\right]$. The non-negative weight is
$W(\sigma)=\zeta(\sigma)\exp (-\sum\vert V_\alpha\vert\vert
\sigma_{\alpha n}\vert^2 \Delta\beta)$, where
$\zeta(\sigma)\equiv{\rm Tr}\,U_\sigma$ is the canonical partition
function of the one-body evolution operator $U_\sigma\equiv U_{N_t}
\ldots U_1$, where $U_n=\exp(-\Delta\beta h_n)$, and the one-body
Hamiltonian for the $n^{\rm th}$ time slice is $h_n= \sum_\alpha
(\epsilon^\ast_\alpha+ s_\alpha V_\alpha \sigma_{\alpha n}) \bar{\cal
O}_\alpha+ (\epsilon_\alpha+ s_\alpha V_\alpha \sigma_{\alpha
n}^\ast) {\cal O}_\alpha$, with $s_\alpha=\pm1$ for $V_\alpha<0$ and
$s_\alpha=\pm i$ for $V_\alpha>0$. The ``sign'' is
$\Phi(\sigma)\equiv \zeta(\sigma)/ \vert\zeta(\sigma)\vert$ and
$\langle{\cal O}\rangle_\sigma\equiv {\rm Tr}\,({\cal
O}U_\sigma)/\zeta(\sigma)$. Both $\zeta(\sigma)$ and $\langle{\cal
O}\rangle_\sigma$ can be evaluated in terms of the $N_s\times N_s$
matrix ${\bf U}_\sigma$ that represents the evolution operator
$U_\sigma$ in the single-particle space.

The sign problem arises because the one-body partition function
$\zeta(\sigma)$ is not necessarily positive, so that the Monte Carlo
uncertainty in the denominator of Eq.~(2) (the $W$-weighted average
sign, $\langle\Phi\rangle$) can become comparable to or larger than
$\langle\Phi\rangle$ itself. In most cases $\langle\Phi\rangle$
decreases exponentially with $\beta$ or with the number of time
slices \cite{Linden92}.

An important class of interactions free from the sign problem (i.e.,
$\Phi(\sigma)\equiv1)$ was found in Ref.~\cite{Lang93}. This occurs
when $V_\alpha<0$ for all $\alpha$ in Eq.~(1). In that case,
$s_\alpha=1$ for all $\alpha$, so that both $h_n$ and $U_\sigma$ are
time-reversal invariant. The eigenvectors of ${\bf U}_\sigma$ then
occur as time-reversed pairs with complex conjugate eigenvalues
$\lambda_i,\lambda_i^\ast$ ($i=1,\ldots,N_s/2$), the grand canonical
partition function $\zeta(\sigma)=\Pi_i\vert1+\lambda_i\vert^2$ is
positive definite, and the canonical partition function for even
$N_v$ is also positive definite.

Based on the above observation, it is possible to decompose $H$ into
its ``good'' and ``bad'' parts, $H=H_G+H_B$, with
\begin{eqnarray}
H_G&=&\sum_\alpha(\epsilon_\alpha^\ast\bar{\cal
O}_\alpha+\epsilon_\alpha {\cal O}_\alpha)+
{1\over2}\sum_{V_\alpha<0}V_\alpha \left\{{\cal O}_\alpha, \bar{\cal
O}_\alpha\right\}\nonumber\\
H_B&=&{1\over2}\sum_{V_\alpha>0}V_\alpha \left\{{\cal O}_\alpha,
\bar{\cal O}_\alpha\right\}\;.
\end{eqnarray}
The ``good'' Hamiltonian $H_G$ includes, in addition to the one-body
terms, all the two-body interactions with $V_\alpha<0$, while the
``bad'' Hamiltonian $H_B$ contains all interactions with
$V_\alpha>0$. By construction, calculations with $H_G$ alone have
$\Phi(\sigma)\equiv1$ and are thus free of the sign problem.

We define a family of Hamiltonians $H_g$ that depend on a continuous
real parameter~$g$ as $H_g=H_G+g H_B$, so that $H_{g=1}=H$. If the
$V_\alpha$ that are large in magnitude are ``good'', we expect that
$H_{g=0}=H_G$ is a reasonable starting point for the calculation of
an observable $\langle{\cal O}\rangle$. One might then hope to
calculate $\langle{\cal O}\rangle_g={\rm Tr}\,({\cal O}e^{-\beta
H_g})/{\rm Tr}\,(e^{-\beta H_g})$ for small $g>0$ and then to
extrapolate to $g=1$, but typically $\langle\Phi\rangle$ collapses
even for small positive $g$. However, it is evident from our
construction that $H_g$ is characterized by $\Phi(\sigma)\equiv1$ for
any $g\leq 0$, since all the ``bad'' $V_\alpha(>0)$ are replaced by
``good'' $g V_\alpha<0$. We can therefore calculate $\langle{\cal
O}\rangle_g$ for any $g\leq0$ by a Monte Carlo sampling that is free
of the sign problem. If $\langle{\cal O}\rangle_g$ is a smooth
function of $g$, it should then be possible to extrapolate to $g=1$
(i.e., to the original Hamiltonian) from $g\leq0$. We emphasize that
$g=0$ is not expected to be a singular point of $\langle{\cal
O}\rangle_g$; it is special only in the Monte Carlo evaluation.

In the nuclear shell model, the two-body interaction can be written
in a density decomposition as \cite{Lang93}
\[
{1\over2}\sum_{abcd}\sum_{KT\pi}
E_{KT}^\pi(ac,bd)
\sum_M(-)^M \rho_{KMT}(ac)\rho_{K-MT}(bd)\;.
\]
Here $\rho_{KMT}=\rho^p_{KM}+(-)^T \rho^n_{KM}$ $(T=0,1)$,
$\rho_{KM}^{(p,n)}(ac)=(a^\dagger_a\times\tilde a_c)_{KM}$ is the
one-body density operator for the pair of proton or neutron orbits
$(a,c)$ coupled to angular momentum $K$ and its $z$-projection $M$,
and $\pi=(-)^{l_a+l_c}=(-)^{l_b+l_d}$ is the parity. The matrices
$E^\pi_{KT}$ are constructed from the two-body matrix elements
$V^\pi_{JT}(ab,cd)$ of good angular momentum $J$, isospin $T$, and
parity $\pi$ through a Pandya transformation. For interactions that
are time-reversal invariant and conserve parity, the
$E^\pi_{KT}(i,j)$ are real symmetric matrices that can be
diagonalized by a real orthogonal transformation. The eigenvectors
$\rho_{KM}(\alpha)$ play the role of ${\cal O}_\alpha$ in Eq.~(1),
and the eigenvalues $\lambda_{K\pi}(\alpha)$ are proportional to
$V_\alpha$. In the Condon-Shortley \cite{Condon35} convention
$\bar{\rho}_{KM}=\pi(-)^{K+M}\rho_{K-M}$ so that the ``good''
eigenvalues satisfy ${\rm
sign}~[\lambda_{K\pi}(\alpha)]=\pi(-)^{K+1}$ \cite{signrule}. To
minimize the number of auxiliary fields required, we use the freedom
to add an arbitrary symmetric interaction to $H$ \cite{Lang93} and
choose $V^S_{JT=1}=V^A_{JT=0}$ so that $E_{KT=1}\equiv 0$. $E_{KT=0}$
is then uniquely determined by the anti-symmetric part of the
interaction through the combination
$\left(V^A_{JT=0}+V^A_{JT=1}\right)$.

To demonstrate the viability and utility of the method, we have
applied it to the mid-$fp$ shell nucleus ${}^{54}$Fe using the
realistic Brown-Richter interaction \cite{Richter91}. The number of
$m$-scheme Slater determinants describing the 6 valence protons and 8
valence neutrons moving among the $N_s=20$ single-particle states of
the $0f_{7/2,5/2}$ and $0p_{3/2,1/2}$ orbitals is
$\left({20\atop6}\right) \left({20\atop8}\right) \approx5\cdot10^9$.
For comparison, the largest model space treated by standard
diagonalization techniques is currently ${}^{48}$Ti \cite{Caurier91}
where the $m$-scheme dimension is $\approx7\cdot10^6$.

Figure~1 (upper) shows the eigenvalues $V_{K\pi\alpha}= \pi(-)^{K}
\lambda_{K\pi} (\alpha)$ of the Brown-Richter interaction; only about
half of the eigenvalues are negative. However, those with the largest
magnitude are all ``good''. It is possible to use an inverse Pandya
transformation to calculate the usual two-body matrix elements
$V^\pi_{JT}(ab,cd)$ for the ``good'' and ``bad'' interactions,
allowing the matrix elements of $H_G$ to be compared in Fig.~1
(lower) with those of the full interaction. The greatest deviation is
for $J=0,T=1$ (the monopole pairing interaction), where $H_G$ is
about twice as attractive as the physical $H$. In all other channels,
$H_G$ and $H$ are quite similar.

We have performed Monte Carlo calculations for $\beta=2~{\rm
MeV}^{-1}$ using $N_t=32$ (so that $\Delta\beta=0.0625~{\rm
MeV}^{-1}$). For $g=-1,-0.8,-0.6,-0.4,-0.2$, and 0, we took
approximately 3300 uncorrelated samples. The computations were
performed on the Intel Touchstone DELTA 512-node parallel computer,
where each node is an Intel i860 processor. Each node produced and
analyzed a sample in about 4 minutes, so that each value of $g$ took
about 25 minutes in total. Selected calculations for larger values of
$\beta$ or $N_t$ show that we have converged to the true ground-state
properties.

The results for various observables are shown in Fig.~2. The
extrapolations to the physical Hamiltonian $(g=1)$ are done by
least-squares polynomials. For each observable except $\langle
H\rangle$, the degree of the polynomial is chosen to be the lowest
for which $\chi^2$ per degree of freedom is less than 1; linear or
quadratic extrapolations are almost always sufficient. For $\langle
H\rangle$, the variational principle implies the additional
constraint of vanishing derivative at $g=1$, in which case a
quadratic or cubic polynomial is used. We have also calculated
response functions $R(\tau)=\langle {\cal O}^\dagger(\tau) {\cal
O}(0)\rangle$ by polynomial extrapolation of our calculations of
$\ln[R_g(\tau)/R_g(0)]$ for $g\leq0$. Fitting $\ln[R_1(\tau)/R_1(0)]$
to a polynomial in $\tau$ allows us to determine moments of the
normalized strength function $f_{\cal O}(E)$, such as $\bar E\equiv
\int E f_{\cal O} (E)dE$. Our overall method was checked in detail
\cite{Alhassid} by comparison with direct diagonalization in the
$sd$-shell using the Brown-Wildenthal interaction \cite{Brown88} and
in the lower $fp$-shell (${}^{44}$Ti) using the Brown-Richter
interaction \cite{Richter91}.

Table 1 summarizes the extrapolated results for various observables.
Note that the statistical uncertainty in these values is proportional
to the uncertainties in the Monte Carlo results for $g\leq0$, and so
can be reduced by increasing the number of samples. The calculated
first moment of the isoscalar quadrupole strength function,
$(1.25\pm0.16)$~MeV, should be compared with the empirical excitation
energy of the first $2^+$ state, 1.408~MeV. Our estimate for the
$B(E2)$ for the decay of this state assuming free nucleon charges
(and that this transition has all of the strength) is $(96\pm1)~{\rm
e^2~fm^4}$, while effective charges $(e_p,e_n)=(1.1,0.1)e$ would be
required to reproduce the experimental value of $126~{\rm e^2fm^4}$.
These charges are significantly smaller than the (1.35, 0.35)$e$ used
in truncated calculations \cite{Auerbach} or the (1.33, 0.64)$e$ used
in the lower $fp$ shell \cite{Richter91}. The total mass quadrupole
strength, $\langle Q^2\rangle= (1482\pm84)~{\rm fm^4}$, is
significantly larger than the simple single particle estimate of
$380~{\rm fm^4}$. The total M1 strength $\langle(M1)^2\rangle=
(14.1\pm0.4)~\mu^2_N$ is quenched relative to the single particle
estimate of $42.55~\mu^2_N$. It is also interesting to note that the
occupation numbers of the single particle orbits are smeared across
the Fermi surface.

Of particular physical interest are the Gamow-Teller operators. Our
calculations exactly satisfy the sum rule $\langle(GT_-)^2\rangle-
\langle(GT_+)^2\rangle= 3(N-Z)=6$. The single particle estimate for
$\langle(GT_+)^2\rangle$ corresponding to the $f_{7/2}~{\rm
proton}\rightarrow f_{5/2}~{\rm neutron}$ transition is 10.28
\cite{Aufderheide}, so the shell model Monte Carlo value of
$4.32\pm0.24$ is quenched by 58\%. This value is comparable to the
experimental result of $3.1\pm0.6$ \cite{Vetterli89}, but
significantly smaller than previous estimates of 6.40 or 6.70 based
on truncated bases \cite{Auerbach}. The additional quenching on the
full space correlates with the enhanced $B(E2, 2^+_1,\rightarrow
0^+_1)$, (i.e., smaller effective charge), as was surmised in
\cite{Auerbach}.

Direct comparison with experimental Gamow-Teller strength functions
requires that we know the energy of the daughter ground state
relative to ${}^{54}$Fe.
Since the ${}^{54}$Co ground state is the isobaric analog state (IAS)
of the ${}^{54}$Fe ground state, we find a mean $(p,n)$ excitation
energy of $\bar E_x=(6.13\pm0.17)$~MeV. This is in agreement with the
systematics of Nakayama {\it et al.} \cite{Nakayama}, which give
$E_{{\rm GT}^-}-E_{\rm IAS}=5.81$~MeV, but is somewhat low relative
to the experimental value of 8.2~MeV \cite{Vetterli89}. When our
calculations of the mean $(n,p)$ excitation energy are corrected for
the Coulomb energy (including exchange) and the nucleon mass
difference, we find $\bar E_x=(1.24\pm0.2)$~MeV, to be compared with
the experimental centroid of 3~MeV \cite{Vetterli89}. A more
consistent theoretical value of $\bar E_x$ can be obtained by
calculating the mass differences of the $A=54$ isobars within the
shell model Monte Carlo \cite{Alhassid}.

The method presented in this Letter is a practical solution to the
sign problem for realistic shell model interactions. A full-basis
calculation of ${}^{54}$Fe with the Brown-Richter interaction shows
the feasibility of the method, with significant quenching of the
Gamow-Teller $\beta^+$ strength. Systematic studies of the
temperature, nuclide, and interaction dependence of these
calculations will be reported elsewhere. Our techniques also enable
the determination of an optimal effective interaction and effective
operators in a greatly enlarged model space.

This work was supported in part by the NSF, Grants No. PHY90-13248
and PHY91-15574, by the DOE, Grant No. DE-FG02-91ER40608, and by a
Caltech DuBridge postdoctoral fellowship to WEO. We thank P.~Vogel
and B.~A.~Brown for helpful discussions, and the Concurrent
Supercomputing Consortium for a grant of DELTA time.

\begin{figure}
\caption{Upper: The eigenvalues $V_\alpha$ of the Brown-Richter
interaction in the $fp$-shell. Eigenvalues for each particle-hole
angular momentum $K$ are plotted in increasing order.
Bottom: The two-body matrix elements $V_{JT=1}(ab,cd)$ of the
Brown-Richter interaction (solid circles) and its ``good'' part (open
circles), for $J\leq4$. The ordering for each $J$ is arbitary. Plots
of the $V_{JT=0}$ and the remaining $T=1$ matrix elements (not shown)
are similar to those shown for $J\geq1$.}
\end{figure}

\begin{figure}
\caption{The results of the Monte Carlo calculations for ${}^{54}$Fe
at $\beta=2~{\rm MeV}^{-1}$ for several observables as a function of
$g\leq0$. $Q=Q_p+Q_n$ is the isoscalar quadrupole, $Q_v=Q_p-Q_n$ is
the isovector quadrupole, $GT_+$ is the Gamow-Teller operator
changing a proton to a neutron, and M1 is the magnetic moment
operator using the free-nucleon $g$-factors. The lines are polynomial
extrapolations; the extrapolated values and corresponding
uncertainties are shown at $g=1$. The extrapolation is linear for
$\langle M1^2\rangle$, but quadratic for $\langle Q^2\rangle$,
$\langle Q^2_v\rangle$, and $\langle{\rm GT}^2_+\rangle$. For
$\langle H\rangle$, the extrapolation is cubic with the constraint of
vanishing derivative at $g=1$.}
\end{figure}

\widetext
\begin{table}
\caption{Monte Carlo results for ${}^{54}$Fe.}
\begin{tabular}{l r@{}l r@{}l}
\multicolumn{5}{c}{$\langle H\rangle=-55.5\pm0.5$~MeV}\\
&\multicolumn{2}{c}{\bf Total strength}&
\multicolumn{2}{c}{\bbox{\bar E~({\rm MeV})}}\\
Isoscalar quadrupole&
$\langle Q^2\rangle$&$\,=\,1482\pm84~{\rm fm^4}$&
1.25&$\;\pm\;0.16$\\
Isovector quadrupole&
$\langle Q_v^2\rangle$&$\,=\,381.3\pm33.8~{\rm fm^4}$&
12.7&$\;\pm\;0.2$\\
Gamow-Teller ($p,n$)&
$\langle(GT_-)^2\rangle$&$\,=\,10.32\pm0.24$&
6.13&$\;\pm\;0.17$\\
Gamow-Teller ($n,p$)&
$\langle(GT_+)^2\rangle$&$\,=\,4.32\pm0.24$&
9.7&$\;\pm\;0.2$\\
$M1$&
$\langle(M1)^2\rangle$&$\,=\,14.1\pm0.4~\mu^2_N$&
8.6&$\;\pm\;0.7$\\
\tableline
\multicolumn{5}{c}{\bf Occupation Numbers}\\
\multicolumn{2}{c}{\bf Protons}& \multicolumn{2}{c}{\bf Neutrons}\\
\multicolumn{2}{c}{$\langle a^\dagger
a\rangle_{f_{7/2}}=4.92\pm0.03$}&
\multicolumn{2}{c}{$\langle a^\dagger
a\rangle_{f_{7/2}}=6.35\pm0.03$}\\
\multicolumn{2}{c}{$\langle a^\dagger
a\rangle_{p_{3/2}}=0.56\pm0.02$}&
\multicolumn{2}{c}{$\langle a^\dagger
a\rangle_{p_{3/2}}=0.86\pm0.02$}\\
\multicolumn{2}{c}{$\langle a^\dagger
a\rangle_{p_{1/2}}=0.11\pm0.01$}&
\multicolumn{2}{c}{$\langle a^\dagger
a\rangle_{p_{1/2}}=0.17\pm0.01$}\\
\multicolumn{2}{c}{$\langle a^\dagger
a\rangle_{f_{5/2}}=0.41\pm0.01$}&
\multicolumn{2}{c}{$\langle a^\dagger
a\rangle_{f_{5/2}}=0.61\pm0.01$}\\
\end{tabular}
\end{table}


\begin{references}
\bibitem{signrule}If all single-particle orbits in the model space
have the same parity, $\pi=+1$ and the sign rule of
Ref.~\protect\cite{Lang93} is reproduced. The sign rule of
Ref.~\protect\cite{Lang93} holds in the Biedenharn-Rose convention,
but in the calculation of the two-body matrix elements the spherical
harmonics $Y_{lm}$ should be multiplied by~$i^l$.
\end{references}
\end{document}